\documentclass[12pt]{iopart}
\usepackage{iopams}
\usepackage{graphicx}
\begin{document}
\paper[Exact solution of the zero-range process]{Exact solution of the zero-range process: fundamental diagram of the corresponding exclusion process}
\author{Masahiro Kanai}%
\address{Graduate School of Mathematical Sciences, The University of Tokyo, 3-8-1 Komaba, Tokyo 153-8914, Japan}
\ead{kanai@ms.u-tokyo.ac.jp}
\begin{abstract}
In this paper, we propose a general way of computing
 expectation values in the zero-range process,
 using an exact form of the partition function.
As an example, we provide the fundamental diagram
 (the flux-density plot) of the asymmetric exclusion
 process corresponding to the zero-range process.
We express the partition function for the steady state by
 the Lauricella hypergeometric function, and thereby have
 two exact fundamental diagrams each for the parallel and
 random sequential update rules.
Meanwhile, from the viewpoint of equilibrium statistical
 mechanics, we work within the canonical ensemble but
 the result obtained is certainly in agreement with previous
 works done in the grand canonical ensemble.
\end{abstract}
\section{Introduction}
{\it The zero-range process} (ZRP) is a stochastic model in
 which many indistinguishable particles occupy sites on a
 lattice \cite{Spitzer70,Evans00,EH05}.
Each site of the lattice may contain an integer number of
 particles and the particles hop to the next site with
 a probability which depends on the number of particles
 at the departure site.
In other words, each particle interacts with the particles
 at the same site, i.e., they have a zero-range interaction.
The hop probability is totally asymmetric, i.e.,
 particles can only move in the definite direction and hence
 the flow of particles, even from the macroscopic viewpoint,
 never vanishes in general.
The flow gives rise to a contrasting density distribution of
 particles, and one may therefore regard the ZRP as a typical
 nonequilibrium system \cite{SZ95,Schutz03}.
Since one can choose any function as the hop probability function,
 the ZRP has been extensively studied and applied
 for a wide variety of many-particle systems \cite{BBJ97,OEC98,KJNSC04}.

Although there now exist a lot of generalized/extended versions
 of the ZRP \cite{EH05,Schutz03,LMS05}, in this paper we focus
 exclusively on the original one, i.e., the ZRP
 in one dimension and with periodic boundary.
Updates of the particle configuration occur at each discrete
 time step, and two typical update rules, the parallel and
 random sequential update rules, are considered.
These rules are defined as follows.
{\it Parallel update rule}: at each time step, simultaneously
 at every site, one particle attempts to hop to the next site
 with its hop probability.
{\it Random sequential update rule}: at each time step, a site is selected
 at random and then one particle at the site attempts to hop to
 the next site with its hop probability.

It is notable that the ZRP in one dimension can be mapped onto
 an exclusion process, i.e., a many-particle system on a
 periodic lattice whose sites contain a single particle or
 none \cite{EH05}.
Hereafter we call particles in the exclusion process
 {\it vehicle} to prevent confusion.
According to this mapping, particles in the ZRP are
 regarded as the distance between adjacent vehicles in the
 exclusion process, i.e., the number of particles at site $i$
 indicates the number of blank sites in front of vehicle $i$
 following vehicle $i-1$; accordingly, for a particle to hop to
 the next site in the ZRP is for a vehicle to hop to the next
 site in the exclusion process.
This exclusion process adequately simulates a traffic flow in
 which vehicles hop forward each with its own probability
 depending on the distance to the front vehicle.
Note that the number of sites in the exclusion process is equal
 to the sum of sites and particles in the ZRP and meanwhile
 the number of vehicles is equal to that of sites in the ZRP.

%
\section{Nonequilibrium steady state and partition function}
In this section, following \cite{EH05}, we precisely define
 the nonequilibrium steady state of the ZRP and then formulate
 the partition function in that steady state.
As well as equilibrium statistical mechanics, expectation
 values of physical quantities should be given
 by using that partition function.
\subsection{Nonequilibrium steady state}
Let $P(\{n_m\})$ be the probability of finding the system in
 a configuration $\{n_m\}=\{n_1,\,n_2,\,\ldots,\,n_M\}$, where
 $n_m$ denotes the number of particles at the $m$th site.
The transition probability from $\{n'_m\}$ to $\{n_m\}$, denoted by
 $T(\{n_m\}|\{n'_m\})$, is expressed by the hop probability function
 $u(n)$ respectively according to the update rule, i.e.,
 parallel and random sequential update rules.
Note that it is always true that $u(0)=0$.

We firstly consider the parallel update rule.
The transition probability for the update rule is given by
\begin{equation}
T(\{n_m\}|\{n'_m\})=\sum^1_{\nu_1=0}\cdots\sum^1_{\nu_M=0}\left[\prod^M_{m=1}u(n'_m)\delta(n_m-n'_m+\nu_m-\nu_{m-1})\right],
\end{equation}
 where, at each site, $\nu_m~(=0,1)$ particle hops to the next
 site with probability $u(n_m)$, and $\delta(n)$ is the
 Kronecker delta that returns unity if $n=0$ and zero otherwise.
The balance of probability currents at each configuration
 $\{n_m\}$ is represented by the following equation:
\begin{equation}
\sum_{\{n'_m\}}\Bigl[T(\{n_m\}|\{n'_m\})P(\{n'_m\})-T(\{n'_m\}|\{n_m\})P(\{n_m\})\Bigr]=0.\label{sscp}
\end{equation}
Then, we call a solution of (\ref{sscp}) {\it the nonequilibrium
 steady state probability} (or the steady state probability,
 simply).
The steady state probability $P(\{n_m\})$ is given as a product
 of the single-site weights $f(n)$:
\begin{equation}
P(\{n_m\})=\frac{1}{Z_{M,N}}\prod^M_{m=1}f(n_m)
\qquad(n_1+n_2+\cdots+n_M=N),
\label{P}
\end{equation}
 where a normalization $Z_{M,N}$, the sum of the products over
 all configurations, is referred to as {\it the partition
 function} for the ZRP in the nonequilibrium steady state.
One can directly confirm that (\ref{P}) is a solution of
 (\ref{sscp}); meanwhile the single-site weight is found to be
\begin{equation}
f(n)=\left\{
\begin{array}{ll}
1-u(1) & (n=0)\\
\displaystyle\frac{1-u(1)}{1-u(n)}\prod^n_{j=1}
\frac{1-u(j)}{u(j)} & (n\geq1).
\end{array}
\right.
\label{fp}
\end{equation}

Next, we consider the random sequential update rule.
In this case, one can describe the steady state condition
 in a simple form:
\begin{equation}
\sum^M_{m=1}\left[u(n_{m-1}+1)P(\,\ldots,\,n_{m-1}+1,\,n_m-1,\,\ldots)
-u(n_m)P(\{n_m\})\right]\theta(n_m),\label{sscr}
\end{equation}
 where $\theta(n_m)$ is the Heaviside function, which emphasizes
 that site $m$ must be occupied for there to be associated hops
 out of and into the configuration $\{n_m\}$.
As well as in the case of the parallel update rule, the steady
 state probability $P(\{n_m\})$ is given in the product form
 (\ref{P}), where the single-site weight for the random
 sequential update rule is
\begin{equation}
f(n)=\left\{
\begin{array}{ll}
1&(n=0)\\
\displaystyle \prod^n_{j=1}\frac1{u(j)}\qquad&(n\geq1).
\end{array}
\right.\label{fr}
\end{equation}
(See \cite{EH05} and \cite{Evans97} for details.)
\subsection{Partition function}
It should be noted that $f(n)$ is not identical to the
 probability that a given site (e.g. site 1) contains $n$
 particles.
Let the probability denoted by $p(n)$, and it is obtained from
 the probability distribution of configurations $P(\{n_m\})$ as
\begin{equation}
p(n)=\sum_{n_2+n_3+\cdots+n_M=N-n}P(\{n,\,n_2,\,\ldots,\,n_M\})
= f(n)\frac{Z_{M-1,N-n}}{Z_{M,N}}.
\label{p}
\end{equation}
The sum of $p(n)$ over $n$ is unity by definition, and we
 thereby obtain the recursion formula for the partition
 functions:
\begin{eqnarray}
Z_{M,N}&=&\sum^{N}_{n=0}f(n)Z_{M-1,N-n}\qquad(M>1,~N\geq0),\label{rec}\\
Z_{1,k}&=&f(k)\qquad(k\geq1).
\end{eqnarray}
Considering the generating functions,
 $\widehat{f}(\zeta):=\sum^\infty_{n=0}f(n)\zeta^n$
 and $\widehat{Z}_M(\zeta):=\sum^\infty_{n=0}Z_{M,n}\zeta^n$,
 we have a recursion with respect only to $M$:
 $\widehat{Z}_{M}(\zeta)=\widehat{f}(\zeta)\widehat{Z}_{M-1}(\zeta)$.
Consequently, we find the fundamental relation between the two generating functions:
\begin{equation}
\widehat{Z}_M(\zeta)=\left(\widehat{f}(\zeta)\right)^M.
\label{genZgenf}
\end{equation}
Thus, the partition function $Z_{M,N}$ is obtained from the
 single-site weight $f(n)$ via their generating functions.
From the statistical mechanical viewpoint, $\widehat{Z}_{M}(\zeta)$ is appropriate for us to call {\it the grand-canonical partition function}, where we should think of $\zeta$ as the fugacity \cite{Evans00,EH05}.

Using $p(n)$ given in (\ref{p}), the average velocity of vehicles in the steady state is defined by
\begin{equation}
v_{M,N}=\sum^N_{n=0}u(n)p(n).
\end{equation}
In the case of parallel update rule, we find that the average
 velocity is expressed by the partition function as
\begin{equation}
v_{M,N}=-\frac{\sum^{N-1}_{n=0}(-1)^nZ_{M,n}}{(-1)^NZ_{M,N}}.
\label{vp}
\end{equation}
In the case of random sequential update rule, we also find
\begin{equation}
v_{M,N}=\frac{Z_{M,N-1}}{Z_{M,N}}.
\label{vr}
\end{equation}
(See \cite{KNT06b} for details.)
\section{Exact solution}
In this section, we exactly compute the partition function and then the expectation value of the average velocity.
These computations are fully done by using the formulae for the
 Lauricella hypergeometric function.
The definition and formulae used are relegated to Appendix.
\subsection{Parallel update rule}
We start with the hop probability function $u(n)$ given as
\begin{equation}
\left\{
\begin{array}{ll}
u(0)=0,&\\
0<u(n)<1&\qquad(1\leq \forall n\leq K),\\
u(n)=1&\qquad(\forall n\geq K+1),
\end{array}
\right.\label{up}
\end{equation}
 where $K~(\geq1)$ is a constant integer.
This is not a strong restriction, for one may take $K$ to be as
 large as necessary.
As far as traffic-flow models are concerned, it is quite
 natural to choose this kind of hop probability function.

Substitution of (\ref{up}) into (\ref{fp}) yields $f(n)=0$ for
 $n>K+1$, and we accordingly transform $\widehat{f}(\zeta)$
 into a convenient form, i.e.,
\begin{equation}
\widehat{f}(\zeta)=\sum^{K+1}_{n=0}f(n)\zeta^n
=\left(1-u(1)\right)(1+\zeta)\sum^{K}_{n=0}\left[\zeta^n\prod^n_{j=1}\frac{1-u(j)}{u(j)}\right],\label{gfp}
\end{equation}
where we define $\prod^n_{j=1}a_j=1$ if $n=0$.
For further computations, we formally factor the most right hand side
 of (\ref{gfp}):
\begin{equation}
\widehat{f}(\zeta)=f(0)(1+\zeta)\prod^K_{i=1}(1+x_i\zeta).\label{fxp}
\end{equation}
Then, comparing (\ref{gfp}) with (\ref{fxp}), we note the
 relation between $x_1,\,x_2,\,\ldots,\,x_K$ and
 $u(1),\,u(2),\,\ldots,\,u(K)$, namely
\begin{equation}
e_0(x)=1,\qquad e_n(x)=\prod^n_{j=1}\frac{1-u(j)}{u(j)}\qquad(n=1,\,2,\,\ldots,\,K),
\label{sym}
\end{equation}
where $e_n(x)$ is the $n$th elementary symmetric function of
 $K$ variables, $x=(x_1,\,\ldots,\,x_K)$.
(Note that the elementary symmetric functions are defined by
 its generating function:
 $\prod_{i}(1+x_i\zeta)=\sum_{n}e_n(x)\zeta^n$,
 and specially $e_0(x)=1$.)

Accordingly, substitution of (\ref{fxp}) into (\ref{genZgenf})
 yields
\begin{equation}
\widehat{Z}_M(\zeta)={f(0)}^M(1+\zeta)^M\prod^K_{i=1}(1+x_i\zeta)^M.\label{genZgenf2}
\end{equation}
Compare (\ref{genZgenf2}) with (\ref{genL}), the generating
 function for the Lauricella hypergeometric function
 $F_{\sf D}$, and one finds an expression for the partition
 function in the nonequilibrium steady state: if $N\leq KM+M$
 then
\begin{equation}
Z_{M,N}={f(0)}^M(-1)^N\frac{(-M)_N}{(1)_N}F_{\sf D}(-N,\,\overbrace{-M,\,\ldots,\,-M}^{K},\,M-N+1;\,x),\label{Zp}
\end{equation}
and otherwise $Z_{M,N}=0$.
(See \ref{app} for definition of the notation.)
As seen below, this expression using the Lauricella
 hypergeometric function is quite useful for exact computation of expectation values.

In the case that $N> KM+M$ the average velocity in the steady
 state takes the value of unity, since all vehicles take more
 than $K$-site distance from the front vehicle (or all sites
 in the corresponding ZRP contain more than $K$ particles).
If $N\leq KM+M$, by substituting of (\ref{Zp}) into (\ref{vp})
 we have an expression for the average velocity.
Using a recursion formula for the Lauricella hypergeometric
 function with respect to parameters (\ref{Lrec}),
\begin{equation}
\alpha F_{\sf D}(\alpha+1,\,\beta,\,\gamma+1;\,x)-\gamma F_{\sf D}(\alpha,\,\beta,\,\gamma;\,x)=(\alpha-\gamma)F_{\sf D}(\alpha,\,\beta,\,\gamma+1;\,x),
\end{equation}
the sum in the numerator in (\ref{vp}) is firstly carried
 out:
\begin{eqnarray}
&\sum^{N-1}_{n=0}\frac{(-M)_n}{(1)_n}F_{\sf D}(-n,\,-M,\,\ldots,\,-M,\,M-n+1;\,x)\nonumber\\
&\qquad\qquad=\frac{(1-M)_{N-1}}{(1)_{N-1}}F_{\sf D}(1-N,\,-M,\,\ldots,\,-M,\,M-N+1;\,x).
\end{eqnarray}
Accordingly, the average velocity of vehicles for arbitrary
 numbers of sites and vehicles is given as
\begin{equation}
v_{M,N}=\frac{N}{M}\frac{F_{\sf D}(1-N,\,-M,\,\ldots,\,-M,\,M-N+1;\,x)}{F_{\sf D}(-N,\,-M,\,\ldots,\,-M,\,M-N+1;\,x)}.
\label{vpL}
\end{equation}
We remark that (\ref{vpL}) is valid for arbitrary numbers of $M$ and $N$.

Then, we expand (\ref{vpL}) in lattice size $L=M+N$
 with a fixed density of vehicles $\rho=M/L$.
Our strategy to take the thermodynamic limit is to eliminate
 the hypergeometric function from (\ref{vpL}).
The Lauricella hypergeometric differential equation that
 $Y=F_{\sf D}(\alpha,\,\beta,\,\gamma;\,x)$ satisfies is defined by
\begin{equation}
\left[(\gamma-1+\delta)\delta_i-x_i(\alpha+\delta)(\beta_i+\delta_i)\right]Y=0\qquad(i=1,\,2,\,\ldots,\,K),\label{HGDE}
\end{equation}
 where $\delta=\sum_i\delta_i$ and
 $\delta_i=x_i\partial/\partial x_i$.
Using a differentiation formula (\ref{dfL}), we have
\begin{equation}
F_{\sf D}(1-N,\,-M,\,\ldots,\,-M,\,M-N+1;\,x)=y-\frac1N\delta y,
\end{equation}
 where $y=F_{\sf D}(-N,\,-M,\,\ldots,\,-M,\,M-N+1;\,x)$,
 the denominator in (\ref{vpL}), and then substitution of this
 into (\ref{vpL}) yields
\begin{equation}
\delta y=\frac{N}{h}\left(h-v_{M,N}\right)y\qquad(h=\frac NM)\label{dy}
\end{equation}
Meanwhile, we consider the hypergeometric differential equation
 for $y$ (where $\alpha=-N$, $\beta_i=-M$ and $\gamma=M-N+1$)
 and find
\begin{equation}
\delta_iy=\frac{N}{h}\frac{v_{M,N}x_i+\frac{h}{N}(1-x_i)\delta_iv_{M,N}}{1-v_{M,N}+v_{M,N}x_i}y\qquad(i=1,\,2,\,\ldots,\,K).\label{diy}
\end{equation}
Since $\sum_i\delta_iy=\delta y$, equating the sum of the right
 hand side of (\ref{diy}) with that of (\ref{dy}) we
 successfully eliminate $y$:
\begin{equation}
h-v_{M,N}=\sum_{i=1}^K\frac{v_{M,N}x_i+\frac{h}{N}(1-x_i)\delta_iv_{M,N}}{1-v_{M,N}+v_{M,N}x_i}.
\end{equation}
Expand the average velocity as
 $v_{M,N}=v_0+v_1L^{-1}+v_2L^{-2}+\cdots$, and substitution of
 this into (\ref{diy}) yields a series of equations for $v_0,\,v_1,\,v_2,\ldots$.
Consequently, as $N$ tends to infinity, we have
\begin{equation}
h-v=\sum^K_{i=1}\frac{vx_i}{1-v+vx_i},\label{hvxp}
\end{equation}
 where $v~(=v_0)$ is the average velocity in the thermodynamic
 limit.
Solving the series of equations for $v_1,\,v_2,\ldots$
 in turn, one can obtain the higher-order correction terms
 as functions of $v$.
The computations are straightforward, but however
 the expressions of those terms are too long to be included.

In \cite{Evans97}, they already obtained the same result as (\ref{hvxp}) working in the grand-canonical ensemble where the fluctuation of the particle number $N$ is allowed.
Actually, (\ref{hvxp}) is consistent with their result,
\begin{equation}
\langle N\rangle=w\frac{\partial}{\partial w}\log\widehat{Z}_M(w),
\label{hgfp}
\end{equation}
where $\langle N\rangle$ means the expectation value of the particle number in the grand-canonical ensemble.
Since the right hand side of (\ref{hvxp}) is in a symmetric
 form with respect to $x_1,\,\ldots,\,x_K$, one can express it
 using the elementary symmetric functions of them.
Recall $\prod_{i}(1+x_i\zeta)=\sum_{n}e_n(x)\zeta^n$,
 and one finds the identity,
\begin{equation}
\sum^{K}_{i=1}\frac{w x_i}{1+w x_i}=\frac{\sum^{K}_{r=1}re_r(x)w^r}{\sum^{K}_{r=0}e_r(x)w^r}.
\label{id}
\end{equation}
The left hand side of (\ref{id}) is equivalent to the right hand
 side of (\ref{hvxp}) if one lets $w=v/(1-v)$, and thus we find that (\ref{hvxp}) is equivalent to (\ref{hgfp}).

\begin{figure}[bth]
\begin{center}
\includegraphics{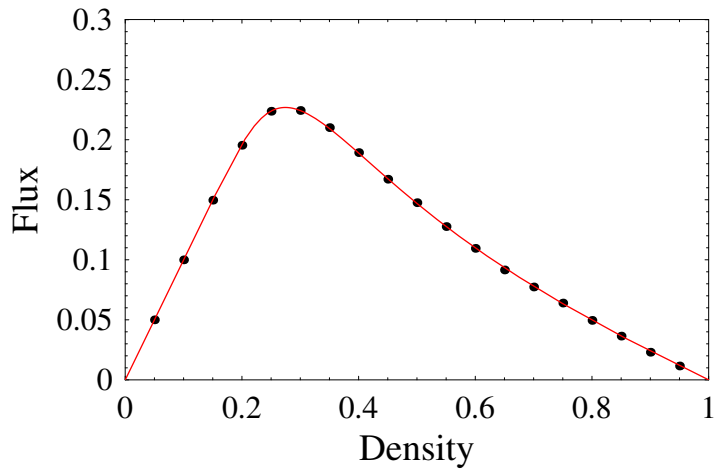}
\caption{
A fundamental diagram of the exclusion process corresponding to
 the zero-range process with the parallel update rule.
Gray line shows the exact solution given in the parametric representation, and black points are simulation data.
Numerical simulation is done with site $L=1000$.
These are in good agreement.
}
\label{fig1}
\end{center}
\end{figure}

We note that using (\ref{hgfp}) one has the fundamental diagram in a parametric representation: flux $Q(v)=\rho(v) v$ and density $\rho(v)=1/(1+h)$, where $v~(0\leq v\leq1)$ is the parameter.
As an example, we illustrate a fundamental diagram choosing a hop probability function that approximately describes motion of vehicles in traffic flow, i.e.,
\begin{equation}
u(n)=\frac{\tanh(n-c)+\tanh c}{1+\tanh c},\label{tfm}
\end{equation}
 where we let $c=3/2$.
Note that the hop probability function can be estimated from real
 traffic data, and (\ref{tfm}) is often chosen for traffic-flow
 models \cite{KNT05,ov}.
Figure \ref{fig1} shows the exact fundamental diagram
 illustrated by using (\ref{hgfp}) with $K=50$, and the simulation
 data in addition.
They show complete agreement.

\subsection{Random sequential update rule}\label{exr}
In the same way as in the case of the parallel update rule,
 we obtain the fundamental diagram for the ZRP with the random
 sequential update rule.

The single-site weight is given in (\ref{fr}), and then the generating function for these weights is given as
\begin{equation}
\widehat{f}(\zeta)=\sum^\infty_{n=0}\left[\zeta^n\prod^n_{j=1}\frac1{u(j)}\right],\label{gfr}
\end{equation}
where $u(n)$ is given in (\ref{up}).
Then, we can factor (\ref{gfr}) into
 $\widehat{f}(\zeta)=(1-\zeta)^{-1}\prod^K_{i=1}(1-x_i\zeta)$
 and express the partition function by
 the Lauricella hypergeometric function:
\begin{equation}
Z_{M,N}=\frac{(M)_N}{(1)_N}F_{\sf D}(-N,\,\overbrace{-M,\,\ldots,\,-M}^K,\,-M-N+1;\,x),\label{Zr}
\end{equation}
where the independent variables $x=(x_1,\,x_2,\,\ldots,\,x_K)$ satisfy
\begin{equation}
e_0(x)=1,\qquad e_n(x)=\prod^n_{j=1}\frac1{u(j)}\qquad(n=1,\,2,\,\ldots,\,K).
\end{equation}
{From} (\ref{vr}) and (\ref{Zr}), the average velocity for
 the finite numbers, $M$ and $N$, is expressed by the
 Lauricella hypergeometric function:
\begin{equation}
v_{M,N}=\frac{N}{M+N-1}\frac{F_{\sf D}(-N+1,\,-M,\,\ldots,\,-M,\,-M-N+2;\,x)}{F_{\sf D}(-N,\,-M,\,\ldots,\,-M,\,-M-N+1;\,x)}.
\end{equation}
Thus, applying the formulae for the hypergeometric functions
 we find
\begin{equation}
h=\frac{N}{M}=\left(1-\frac1M\right)\frac{v_{M,N}}{1-v_{M,N}}-\sum^K_{i=1}\frac{x_iv_{M,N}+\frac{1-x_i}{(1-v_{M,N})M}\delta_iv_{M,N}}{1-x_iv_{M,N}}.
\end{equation}
As well as the parallel update case, we can expand $v_{M,N}$ as a series $v_0+v_1L^{-1}+v_2L^{-2}+\cdots$.

Especially in the thermodynamic limit, we recover the result given in\cite{BBJ97,EH05}
\begin{equation}
\langle N\rangle=v\frac{\partial}{\partial v}\log\widehat{Z}(v),\label{hgfr}
\end{equation}
which was obtained within the grand-canonical ensemble.

In contrast with the parallel-update case, if $K$ tends to infinity, we need to determine the range of parameter $v$.
For this purpose, we impose a simple condition on the hop function, i.e.,
\begin{equation}
\lim_{n\rightarrow\infty}u(n)=u_{\infty}\qquad(0<u_\infty\leq1),\label{uinf}
\end{equation}
 which surely allows us to determine the parameter range.
Then, the range of values of $v$ becomes $0\leq v<u_\infty$.
It depends on the asymptotic behaviour of $u(n)$ whether
 $\widehat{f}(v)$ converges or not at $v=u_\infty$.
The convergence of $\widehat{f}(v)$ is discussed
 in the context of condensation occurring in the ZRP
 \cite{EH05,MEZ05,EMZ06}.

\begin{figure}[bth]
\begin{center}
\includegraphics{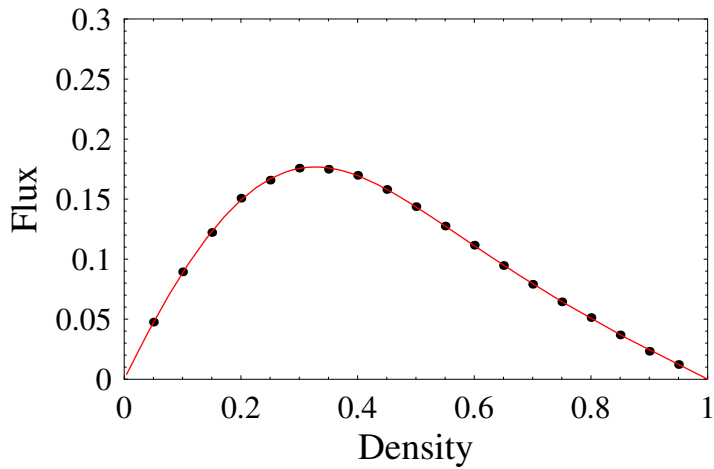}
\caption{
A fundamental diagram of the exclusion process corresponding to the zero-range process with the random sequential update rule.
(See Fig.\,\ref{fig1} for comparison.)
Gray line shows the exact solution given in the parametric representation, and black points are simulation data.
Numerical simulation is done with site $L=1000$.
These are in good agreement.
}
\label{fig2}
\end{center}
\end{figure}

We choose a traffic-flow model with the hop probability function (\ref{tfm}) and random sequential update rule, and show the exact fundamental diagram and simulation result in figure \ref{fig2}.
These are in good agreement as well as in the case of the parallel update rule.

We consider a simple hop probability function which allows us to
 provide a direct formula of fundamental diagram instead of
 the parametric representation, i.e.,
\begin{equation}
u(1)=\lambda,\qquad u(n)=p\quad(n\geq2).\label{Klauck}
\end{equation}
Substitution of (\ref{Klauck}) into (\ref{hgfr}) via (\ref{gfr})
 yields
\begin{equation}
h=\frac{p^2v}{(p-v)(\lambda p+(p-\lambda)v)}.
\end{equation}
Accordingly, eliminating $v$ we obtain
 the fundamental diagram explicitly described by
\begin{equation}
Q=p\rho\left[1-\frac{1-\sqrt{1-4(1-q)\rho(1-\rho)}}{2(1-q)(1-\rho)}\right]\quad\mbox{where}\quad q=\frac\lambda p.\label{FDKS}
\end{equation}
One can also figure out an explicit formula for this model
 in the case of the parallel update rule; however, one has to
 solve a cubic equation.

If we take the limit as $\lambda\rightarrow p$ (i.e.,
 $q\rightarrow1$) in (\ref{FDKS}), we recover the fundamental
 diagram of the ASEP, $Q=p\rho(1-\rho)$, which is a well-known
 solution \cite{RSSS98}.
If we let $p=1$ ($q=\lambda$, accordingly), then we recover
 another exact solution given in \cite{KS99}, where they
 obtained the same result by using the matrix-product ansatz.
\section{Summary and remark}
In this paper, we provide an exact solution of the zero-range
 process:
 we have the partition function for the general case, which is
 expressed by the Lauricella hypergeometric function, and which
 allows us to make further computations.
Although the solution obtained is rather formal, that is enough
 for us to compute expectation values of macroscopic quantities.
In particular, using the exact partition function we obtain
 the average velocity of vehicles for arbitrary numbers of sites and particles.
Moreover, expanding the average velocity in terms of the site number, we obtain the fundamental diagram as the leading term of the expansion.
One can also figure out the correction terms through a long computation.


Due to the fact that the steady state of the ZRP has a
 factorized form, one can do within equilibrium statistical
 mechanics while dealing with the macroscopic quantities
 averaged over all the lattice.
In this regard, one may take $Z_{M,N}$ as the {\it canonical}
 partition function in which the particle number is conserved.
As a result, we compute the expectation value of the average velocity within the canonical ensemble.
In general, statistical treatment in the canonical ensemble is more difficult than that in the grand canonical ensemble.
Both results in these ensembles are equivalent in the thermodynamic limit.
An analytical approach based on the grand canonical ensemble was previously made to the ZRP with the random sequential update rule in \cite{BBJ97}, and to the ZRP with the parallel update rule in \cite{Evans97}.
Recently, canonical analysis of the condensation is also done
 \cite{MEZ05,EMZ06}.
In this regard, it will be interesting to exactly compute
 the probability $p(n)$ that a single site is to contain
 $n$ particles in the steady state.


Recently, as a traffic-flow model, we introduce
 {\it the stochastic optimal velocity (SOV) model}
 \cite{KNT05,KNT06a} in the form of combining two exactly
 solvable models, i.e., the asymmetric simple exclusion process (ASEP) \cite{RSSS98} appearing as a special case of
 the ZRP, and the ZRP itself; however it does not seem to us
 that the model has an exact solution in the general case.
In numerical simulations, we find that the SOV model has
 an intriguing and complex phase transition in the fundamental
 diagram, and our future goal is to analyse in detail the phase
 transition observed, e.g., by using a theoretical method for
 exactly solvable models.
\ack
The author thanks Teruhisa Tsuda for a helpful discussion
 on practical computations.

This work is supported by a grant for the 21st Century COE
 program at The University of Tokyo from the Ministry of
 Education, Culture, Sports, Science and Technology, Japan.
\appendix
\section{The Lauricella hypergeometric function}\label{app}
There are four types of the Lauricella hypergeometric function,
 being denoted by $F_{\sf A}$, $F_{\sf B}$, $F_{\sf C}$ and
 $F_{\sf D}$.
Each of them has a series representation, an integral
 representation and a differential equation it satisfies.
Since only $F_{\sf D}$ appears in this paper, we confine our
 attention to $F_{\sf D}$.
(See \cite{HG} for details.)

Firstly, we define the Lauricella hypergeometric function
 $F_{\sf D}$ with $K$ arguments by
\begin{equation}
F_{\sf D}(\alpha,\,\beta_1,\,\ldots,\,\beta_K,\,\gamma;\,x_1,\,\ldots,\,x_K)=\sum_{m\in I_K}\frac{(\alpha)_{|m|}(\beta)_{m}}{(\gamma)_{|m|}(1)_{m}}x^m,\label{LHG}
\end{equation}
where $\alpha$, $\beta_1,\,\ldots,\,\beta_K$ and $\gamma$ are
 all complex parameters, $(a)_n=a(a+1)\cdots(a+n-1)$ is the
 Pochhammer symbol, and notations in the right hand side are
 defined as follows:
 $I_K=\{m=(m_1,\,\ldots,\,m_K)\,;\,m_i\in\mathbb{Z}_{\geq0}~(i=1,\,2,\,\ldots,\,K)\}$, then, for $m\in I_K$, $|m|=m_1+\cdots+m_K$,
 $(\beta)_m=(\beta_1)_{m_1}(\beta_2)_{m_2}\cdots(\beta_K)_{m_K}$, $(1)_m=(1)_{m_1}(1)_{m_2}\cdots(1)_{m_K}$
 and $x^m={x_1}^{m_1}\cdots{x_K}^{m_K}$.
We also use the following notations for simplicity:
 $\beta=(\beta_1,\,\beta_2,\,\ldots,\,\beta_K)$ and
 $x=(x_1,\,x_2,\,\ldots,\,x_K)$.

The hypergeometric differential equation for $F_{\sf D}(\alpha,\,\beta,\,\gamma;\,x)$ is given in (\ref{HGDE}).
We omit the integral representation because it is not used in the present work.
\subsection{Generating function}
We present the generating function for $F_{\sf D}$:
\begin{equation}
(1-\zeta)^{\kappa}\prod^K_{i=1}(1-x_i\zeta)^{-\beta_i}
=\sum^\infty_{n=0}\frac{(-\kappa)_n}{(1)_n}F_{\sf D}(-n,\,\beta,\,\kappa-n+1;\,x)\zeta^n.
\label{genL}
\end{equation}
Here, we give a proof of (\ref{genL}) by mathematical induction.
One should note the relations,
\begin{equation}
(1-z)^{a}=\sum^\infty_{n=0}\frac{(-a)_n}{(1)_n}z^n,
\end{equation}
and
\begin{equation}
(a)_{l-m}=\frac{a(a+1)\cdots(a+l-1)}{(a+l-m)\cdots(a+l-1)}=\frac{(-1)^m(a)_l}{(1-a-l)_{m}}.
\end{equation}
\begin{enumerate}
\item If $K=1$, then one has
\begin{eqnarray}
\fl
(1-\zeta)^{\kappa}(1-x_1\zeta)^{-\beta_1}
&=&\sum^\infty_{m_1=0}\sum^\infty_{k=0}\frac{(-\kappa)_k(\beta_1)_m}{(1)_k(1)_{m_1}}{x_1}^{m_1}\zeta^{k+m_1}\nonumber\\
&=&\sum^\infty_{m_1=0}\sum^\infty_{n=m_1}\frac{(-\kappa)_{n-m_1}(\beta_1)_{m_1}}{(1)_{n-m_1}(1)_{m_1}}{x_1}^{m_1}\zeta^{n}\nonumber\\
&=&\sum^\infty_{n=0}\frac{(-\kappa)_n}{(1)_n}F_{\sf D}(-n,\,\beta_1,\,\kappa-n+1,\,x_1)\zeta^{n}.
\end{eqnarray}
Thus, (\ref{genL}) holds if $K=1$.
\item Assume that (\ref{genL}) holds if $K=l-1$,
 and one accordingly has
\begin{eqnarray}
\fl
(1-\zeta)^{\kappa}\prod^{l}_{i=1}(1-x_i\zeta)^{-\beta_i}&=&(1-x_{l}\zeta)^{-\beta_l}\sum^\infty_{n=0}\frac{(-\kappa)_n}{(1)_n}F_{\sf D}(-n,\,\beta',\,\kappa-n+1;\,x')\zeta^n\nonumber\\
&=&\sum^\infty_{k=0}\sum^\infty_{n=0}\frac{(-\kappa)_n}{(1)_n}\zeta^{k+n}\sum_{m\in I_{l-1}}\frac{(-n)_{|m|}(\beta')_{m}(\beta_l)_k(x')^{m}{x_l}^k}{(\kappa-n+1)_{|m|}(1)_{m}(1)_k}\nonumber\\
&=&\sum^\infty_{n=0}\frac{(-\kappa)_{n}}{(1)_{n}}\zeta^{n}\sum_{(m,k)\in I_l}\frac{(-n)_{|m|+k}(\beta')_{m}(\beta_l)_k(x')^{m}{x_l}^k}{(\kappa-n+1)_{|m|+k}(1)_{m}(1)_k},
\end{eqnarray}
where $x'=(x_1,\,\ldots,\,x_{l-1})$, $\beta'=(\beta_1,\,\ldots,\,\beta_{l-1})$ and $I_k=\{m=(m_1,\,\ldots,\,m_k)\,;\,m_i\in\mathbb{Z}_{\geq0}~(i=1,\,2,\,\ldots,\,k)\}$.
\end{enumerate}
Thus, (\ref{genL}) holds also when $K=l$.
Therefore, we may conclude that (\ref{genL}) holds for any
 natural number $K$.
\subsection{Recursion formula}\label{Lrec}
We provide a recursion with respect to the parameters,
 $\alpha$ and $\gamma$, for $F_{\sf D}$:
\begin{equation}
\alpha F_{\sf D}(\alpha+1,\,\beta,\,\gamma+1;\,x)=\gamma F_{\sf D}(\alpha,\,\beta,\,\gamma;\,x)+(\alpha-\gamma)F_{\sf D}(\alpha,\,\beta,\,\gamma+1;\,x).\label{recL}
\end{equation}
One should note simple formulae for the Pochhammer symbol,
\begin{equation}
(a+1)_n-(a)_n=\frac{n}a(a)_n,\quad\mbox{and}\quad
\frac1{(a)_n}-\frac1{(a+1)_n}=\frac{n}{(a)_{n+1}}.
\label{aaa}
\end{equation}
{From} (\ref{LHG}) and the two formulae above, one respectively has
\begin{equation}
F_{\sf D}(\alpha+1,\,\beta,\,\gamma;\,x)-F_{\sf D}(\alpha,\,\beta,\,\gamma;\,x)=\sum_{m\in I}\frac{|m|}{\alpha}\frac{(\alpha)_{|m|}(\beta)_{m}}{(\gamma)_{|m|}(1)_{m}}x^m,\label{al}
\end{equation}
and
\begin{equation}
F_{\sf D}(\alpha,\,\beta,\,\gamma;\,x)-F_{\sf D}(\alpha,\,\beta,\,\gamma+1;\,x)=\sum_{m\in I}\frac{|m|}{\gamma}\frac{(\alpha)_{|m|}(\beta)_{m}}{(\gamma+1)_{|m|}(1)_{m}}x^m.\label{gam}
\end{equation}
Then, we can construct (\ref{recL}) from (\ref{al}) and (\ref{gam}).
\subsection{Differentiation formula}
Since the Lauricella hypergeometric function is a multivariable
 function, when taking the derivative one needs to deal with
 partial differentiation: $\delta_i=x_i\partial/\partial x_i$
 and $\delta=\sum_i\delta_i$.
In the present work, we specially use the following formulae:
\begin{eqnarray}
\delta F_{\sf D}(\alpha,\,\beta,\,\gamma;\,x)&=&\alpha\left(F_{\sf D}(\alpha+1,\,\beta,\,\gamma;\,x)-F_{\sf D}(\alpha,\,\beta,\,\gamma;\,x)\right)\nonumber\\
&=&(\gamma-1)\left(F_{\sf D}(\alpha,\,\beta,\,\gamma-1;\,x)-F_{\sf D}(\alpha,\,\beta,\,\gamma;\,x)\right).
\label{dfL}
\end{eqnarray}
Note that $\delta_i {x_i}^{m_i}=m_i{x_i}^{m_i}$, and
 from (\ref{aaa}) one finds
\begin{eqnarray}
\delta F_{\sf D}(\alpha,\,\beta,\,\gamma;\,x)&=&\sum_{m\in I}\frac{(\alpha)_{|m|}(\beta)_{m}}{(\gamma)_{|m|}(1)_{m}}|m|x^m\nonumber\\
&=&\sum_{m\in I}\frac{\alpha\left((\alpha+1)_{|m|}-(\alpha)_{|m|}\right)(\beta)_{m}}{(\gamma)_{|m|}(1)_{m}}x^m\nonumber\\
&=&\sum_{m\in I}\left(\frac{\gamma-1}{(\gamma-1)_{|m|}}-\frac{\gamma-1}{(\gamma)_{|m|}}\right)\frac{(\alpha)_{|m|}(\beta)_{m}}{(1)_{m}}x^m.
\end{eqnarray}
Thus, (\ref{dfL}) is proven.


\section*{References}
\begin{thebibliography}{99}
\bibitem{Spitzer70}
Spitzer F 1970 {\it Advances in Math.} {\bf 5} 246
\bibitem{Evans00}
Evans M R 2000 {\it Braz. J. Phys.} {\bf 30} 42
\bibitem{EH05}
Evans M R and Hanney T 2005 {\it J. Phys. A: Math. Gen.} {\bf 38} R195
\bibitem{SZ95}
Schmittmann B and Zia R P K 1995 {\it Phase Transitions and Critical Phenomena} vol~17 (London: Academic Press)
\bibitem{Schutz03}
Sch{\" u}tz G M 2003 {\it J. Phys. A: Math. Gen.} {\bf 36} R339
\bibitem{BBJ97}
Bialas P, Burda Z and Johnston D 1997 {\it Nuclear Physics B} {\bf 493} 505
\bibitem{OEC98}
O'Loan O J, Evans M R and Cates M E 1998 {\it Phys. Rev. E} {\bf 58} 1404
\bibitem{KJNSC04}
Kunwar A, John A, Nishinari K, Schadschneider A and Chowdhury D 2004 {\it J. Phys. Soc. Jpn.} {\bf 73} 2979
\bibitem{LMS05}
Levine E, Mukamel D and Sch{\"u}tz G M 2005 {\it J. Stat. Phys.} {\bf 120} 759
\bibitem{Evans97}
Evans M R 1997 {\it J. Phys. A: Math. Gen.} {\bf 30} 5669
\bibitem{KNT05}
Kanai M, Nishinari K and Tokihiro T 2005 {\it Phys. Rev. E} {\bf 72} 035102(R)
\bibitem{KNT06b}
Kanai M, Nishinari K and Tokihiro T 2006 {\it J. Phys. A: Math. Gen.} {\bf 39} 9071
\bibitem{ov}
Bando M, Hasebe K, Nakayama A, Shibata A and Sugiyama Y 1995 {\it Phys. Rev. E} {\bf 51} 1035\\
Bando M, Hasebe K, Nakanishi K, Nakayama A, Shibata A and Sugiyama Y 1995 {\it J. Phys. I France} {\bf 5} 1389\\
Igarashi Y, Itoh K and Nakanishi K 1999 {\it J. Phys. Soc. Jpn.} {\bf 68} 791\\
Hasebe K, Nakayama A and Sugiyama Y 1999 {\it Phys. Lett. A} {\bf 259} 135
\bibitem{MEZ05}
Majumdar S N, Evans M R and Zia R K P 2005 {\it Phys. Rev. Lett.} {\bf 94} 180601
\bibitem{EMZ06}
Evans M R, Majumdar S N and Zia R K P 2006 {\it J. Stat. Phys.} {\bf 123} 357
\bibitem{RSSS98}
Rajewsky N, Santen L, Schadschneider A and Schreckenberg M 1998 {\it J. Stat. Phys.} {\bf 92} 151
\bibitem{KS99}
Klauck K and Schadschneider A 1999 {\it Physica A} {\bf 271} 102
\bibitem{KNT06a}
Kanai M, Nishinari K and Tokihiro T 2006 {\it J. Phys. A: Math. Gen.} {\bf 39} 2921
\bibitem{HG}
Erd{\'e}lyi A 1953-55 {\it Higher Transcendental Functions} (MacGraw-Hill)
\end{thebibliography}
\end{document}